\def\be{\begin{equation}}
\def\ee{\end{equation}}
\def\bea{\begin{eqnarray*}}
\def\eea{\end{eqnarray*}}

\documentclass[useAMS,usenatbib]{mn2e}
\usepackage{epsfig}
\usepackage{subfigure}
\usepackage{graphicx}
\usepackage{flafter}
\usepackage{lscape}
\newcommand{\Eq}{{Equation~}}
\newcommand{\beq}{\begin{equation}}
\newcommand{\eeq}{\end{equation}}

\def\lsim{\mathrel{\lower2.5pt\vbox{\lineskip=0pt\baselineskip=0pt
           \hbox{$<$}\hbox{$\sim$}}}}
\def\gsim{\mathrel{\lower2.5pt\vbox{\lineskip=0pt\baselineskip=0pt
           \hbox{$>$}\hbox{$\sim$}}}}

\input epsf

\title[AzTEC mm survey of the COSMOS field - II] 
{AzTEC Millimetre Survey of the COSMOS Field - II. Source Count Overdensity
and Correlations with Large-Scale Structure}
\author[J.E. Austermann et al.]{
J.E.~Austermann,$^1$
I.~Aretxaga,$^2$
D.H.~Hughes,$^2$
Y.~Kang,$^3$ 
S.~Kim,$^3$
J.D.~Lowenthal,$^4$
\newauthor
T.A.~Perera,$^1$
D.B.~Sanders,$^5$
K.S.~Scott,$^1$
N.~Scoville,$^6$
G.W.~Wilson,$^1$ 
M.S.~Yun,$^1$
\\
$^1$Department of Astronomy, University of Massachusetts, Amherst, MA 01003, USA.\\
$^2$Instituto Nacional de Astrof\'{\i}sica, \'Optica y Electr\'onica
(INAOE), Aptdo. Postal 51 y 216, 72000 Puebla, Pue., Mexico.  \\
$^3$Astronomy \& Space Science Department, Sejong University, Seoul, South Korea.\\
$^4$Department of Astronomy, Smith College, Northampton, MA 01063, USA.\\
$^5$Institute for Astronomy, University of Hawaii, 2680 Woodlawn Drive, Honolulu, HI 96822, USA.\\
$^6$California Institute of Technology, Pasadena, CA 91125.\\
}

\begin{document}

\date{}

\pagerange{} \pubyear{}

\maketitle

\label{firstpage}

\begin{abstract}
We report an over-density of bright sub-millimetre galaxies (SMGs) in the 0.15~deg$^2$ 
AzTEC/COSMOS survey  
and a spatial correlation between the SMGs and the 
optical-IR galaxy density at $z\lsim$~1.1.   
This portion of the COSMOS field shows a $\sim3\sigma$ over-density 
of robust SMG detections when compared to
a background, or ``blank-field'', population model 
that is consistent with SMG surveys of fields with no extragalactic bias.
The SMG over-density is most significant in the number of very bright 
detections (14 sources with 
measured fluxes $S_{1.1mm} > 6$~mJy), which is entirely incompatible with 
sample variance within our adopted blank-field number densities 
and infers an over-density significance of $\gg 4\sigma$.   
We find that the over-density and spatial correlation to optical-IR galaxy density
are most consistent 
with lensing of a background SMG population by foreground mass structures along
the line of sight,
rather than  
physical association of the SMGs with the $z\lsim$~1.1 galaxies/clusters.
The SMG positions are only weakly correlated with weak-lensing maps 
, suggesting that the dominant sources of correlation 
are individual galaxies and the more tenuous structures in the region and not 
the massive and compact clusters.
These results highlight the important roles cosmic variance and large-scale structure 
can play in the study of SMGs.
\end{abstract}


\section{Introduction}

Foreground structure, cosmic variance, and source environment can affect the
observer's perception and interpretation of the source population being 
probed in a particular survey.  
For example, gravitational lensing by massive foreground clusters
affects both the observed flux of sources and the 
areal coverage of the survey in the source plane.   
These aspects of gravitational lensing have been utilised to 
probe the very faint sub-millimetre galaxy (SMG) population  
below the confusion limit imposed by the 
high density of faint SMGs relative to the survey beam size
\citep[e.g.][]{smail97,chapman02a,cowie02,smail02,knudsen06,wilson08b}.
The measured (sub)millimetre fluxes of sources found in the direction 
of very massive clusters can also be affected by, and confused with, the 
signal imposed through the Sunyaev-Zel'dovich effect  
on the cosmic microwave background \citep[e.g.][]{wilson08b}. 
Furthermore, surveys  
can be affected by foreground structures with  
high galaxy-densities, which increase
the likelihood of galaxy-galaxy lensing by intervening 
galaxies and complicate counterpart identification at other 
wavelengths \citep{chapman02b,dunlop04}.

Spectroscopic observations have shown that the vast majority of SMGs with 
detectable radio 
counterparts lie at an average redshift of $z\sim$~2.2 \citep{chapman05}, while 
spectroscopic \citep{valiante07} and photometric \citep[e.g.][]{younger07} analysis put many 
radio-faint SMGs at even higher redshifts. 
The average SMG is unlikely to be found at $z\lsim1$, however
it remains to be seen if the $z\sim1$ SMG population can 
be locally enhanced due to large-scale structure and cosmic variance. 
Some evidence exists for increased number densities of SMGs in 
mass-biased regions of the $z \gsim 1$ Universe.
Surveys towards several $z \sim 1$ clusters \citep{best02,webb05}
find a number density of SMGs in excess of the blank-field counts
that can not be explained by gravitational lensing alone. 
This implies that some of the SMGs are physically associated with the
clusters, although the number statistics are small and the lensing 
could be underestimated \citep{webb05}.  
Similar over-densities have been found towards high-redshift radio 
galaxies \citep{stevens03,debreuck04,greve07} and 
$z > 5$ quasars \citep{priddey08}, where lensing of background
sources is less likely to be an issue.
Spectroscopic observations have also found common redshifts 
amongst SMGs in the SSA22 and HDF fields, suggesting physical 
over-densities of SMGs at redshifts of 3.1 and 2.0, respectively
\citep{chapman05}. 
Together, these surveys suggest that these massive dusty starbursts are
prominent in moderate and high-redshift cluster/proto-cluster 
environments.  

In this paper, we analyse the density and distribution of SMGs
in the AzTEC/COSMOS survey \citep{scott08}.  
The AzTEC/COSMOS survey covers a region within the
COSMOS field \citep{scoville07a} known to contain a high
density of optical-IR galaxies and prominent large-scale structure
at $z \lsim 1.1$ \citep{scoville07b}, 
including a massive $M \sim 10^{15} M_\odot$ cluster at $z \approx 0.73$  
\citep{guzzo07}.
In \S~\ref{sec:counts} we present the 1.1~mm source 
counts for the AzTEC/COSMOS field, 
revealing a strong over-density of bright SMGs compared to the blank-field.
We explore the nature of this over-density through examination of the
spatial correlation between SMGs and the known large-scale structures
(\S~\ref{sec:LSScorr}).
Positive correlation between SMG positions and low-redshift 
large-scale structure has been previously detected statistically in 3 disjoint
SCUBA surveys \citep{almaini03,almaini05}.
We now present a wide-field investigation of such correlations 
using the AzTEC/COSMOS survey, which has advantages in 
its contiguous size (0.15~deg$^2$), broad range of low-redshift environments,
and the availability of deep multiband imaging 
and reliable photometric redshifts \citep{ilbert08}.

This is the second paper describing the 1.1~mm results of the 
AzTEC/COSMOS survey.  
Paper I \citep{scott08}
presented the data-reduction algorithms, AzTEC/COSMOS map and source
catalogue, and confirmation of robustness of the AzTEC/JCMT data and pointing.
Additionally, seven of the brightest AzTEC/COSMOS sources have had 
high-resolution follow-up imaging at 890~$\micron$ using the 
Sub-Millimetre Array (SMA) and are discussed in detail in \citet{younger07}.
Spitzer IRAC colours of these SMGs, and others, are discussed in
\citet{yun08}.

\section{Number counts in the COSMOS field}
\label{sec:counts}

The number density of SMGs provides constraints on 
galaxy evolution models 
\citep[e.g.][]{kaviani03,granato04,baugh05,negrello07}
and insights to 
the dust-obscured component of star formation 
in the high-redshift Universe.  
The number density also describes how these 
discrete objects contribute to the cosmic infrared background 
(CIB), as discussed in Paper I.
In this paper, we focus on how the localised SMG number 
counts reflect large-scale structure.  
Before presenting the number counts for the AzTEC/COSMOS
survey (\S~\ref{ssec:countscos} \& \S~\ref{ssec:countsod}),
we describe the technical details of the flux corrections
(\S~\ref{ssec:fluxcorrection}) and methods
(\S~\ref{ssec:countssub}) that are vital to the construction 
of unbiased source counts from typical SMG surveys.
Here we expand on the flux correction techniques of \citet{coppin06}
and provide new tests of these methods through simulation.

\subsection{\label{ssec:fluxcorrection}Flux Corrections}

\begin{figure}
\epsfig{file=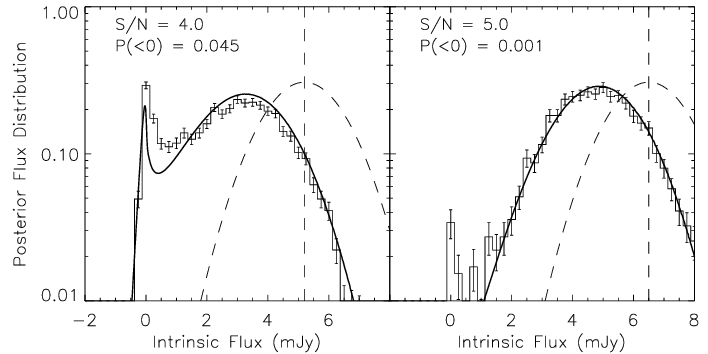, width=1.0\hsize,angle=0}
\caption{Posterior Flux Distribution (PFD)  
  for $S/N=4$ (left) and $S/N=5$ (right) detections in a map with noise 
  $\sim1.3$~mJy assuming an underlying source 
  population consistent with the AzTEC/COSMOS results
  presented in \S~\ref{ssec:countscos}.
  The simulated probability distribution is shown as the histogram
  and is calculated as described in \S~\ref{ssec:fluxcorrection}. 
  Error bars represent the 
  1$\sigma$ Poisson errors of the simulation results, limited only by 
  the number of simulations computed.  
  The Bayesian approximation is depicted as a solid curve.
  The dashed vertical line represents the measured flux, $S_m$. 
  The dashed curve is the Gaussian probability distribution, $p(S_m,\sigma_m|S_i)$, 
  which represents the distribution that might otherwise be 
  assumed without flux
  boosting and/or false detection considerations.
}  
\label{fig:pfd1}
\end{figure}

Surveys of source populations
whose numbers decline with increasing flux result in blind detections that are biased
systematically high in flux.  
This bias is typically referred to 
as ``flux boosting'' 
and results from the fact that detected sources have a higher probability of being an
intrinsically dim source (numerous) coincident with a positive noise fluctuation 
than being a relatively bright source (scarce) coincident with negative noise.  
This effect is concisely described in \citet{hogg98} and is
extremely important for SMG surveys (see Figure~\ref{fig:pfd1})
due to the relatively low $S/N$ of the measurements and 
a population that is known to decline steeply with increasing flux
\citep[e.g.][and references therein]{scott06,coppin06}.

We calculate an intrinsic flux probability
distribution for each potential AzTEC source using the 
Bayesian techniques outlined in Paper I and \citet{coppin05,coppin06}.  
The probability of a 
source having intrinsic flux $S_i$ when discovered in a blind survey with measured
flux $S_m\pm\sigma_m$ is approximated as
\be
\label{eq:bayes}
p(S_i|S_m,\sigma_m) = {p(S_i)p(S_m,\sigma_m|S_i) \over p(S_m,\sigma_m)}
\ee
where $p(S_i)$ is the assumed prior distribution of flux
densities, $p(S_m,\sigma_m|S_i)$ is the likelihood of observing $(S_m,\sigma_m)$
for a source of intrinsic flux $S_i$, 
and $p(S_m,\sigma_m)$ is a normalising constant. 
The resulting probability distribution  
is referred to as the posterior flux distribution, or PFD, throughout this section.
We assume a Gaussian (normal) noise distribution for $p(S_m, \sigma_m|S_i)$
that is consistent with the noise in our map at the location of the discovered source. 
The prior, $p(S_i)$, is generated from pixel histograms of 10,000 
noiseless simulations of the astronomical sky 
-- as would be seen in  
zero-mean AzTEC/JCMT maps -- 
given our best estimate of the true underlying SMG 
population and distribution.  For this paper, we assume the SMG population exhibits number count 
densities that are well described by a Schechter function \citep{schechter76} 
of the form
\be
\label{eq:schechter}
{dN \over dS} = {N^\ast \over S^\prime} \left({S \over S^\prime}\right)^{\alpha+1} \mbox{exp}(-S/S^\prime).
\ee
This parametric form is a slight departure from that used in Paper I and 
in the SCUBA/SHADES survey \citep{coppin06}, with $N^\ast/S^\prime$ replacing the parameter 
$N^\prime$ found in Paper I.  
This form has the advantage of reducing the correlations between the normalising
parameter and the parameters $S^\prime$ and $\alpha$.
The normalising factor, $N^\ast$, is in units of deg$^{-2}$ and is independent of the
observation wavelength when assuming the same source population and a constant
flux ratio between the observing bands.  
For the Bayesian prior, we initially assume parameters of 
$[S^\prime,N^\ast,\alpha] = [1.34,5280,-2]$, which 
represent the best-fit Schechter function to the 
SCUBA/SHADES number counts \citep{coppin06} when 
reparameterised to the form of \Eq~\ref{eq:schechter} and 
scaled to 1.1~mm 
assuming an 850~\micron/1100~\micron~spectral index of 3.5 
(flux ratio $\sim$~2.5).   

A second systematic flux bias in low-$S/N$ blind surveys results from source detections 
being defined as peak locations in the map. 
The measured source flux is, on average, biased high due to the possibility of large positive noise peaks 
lying nearby, but off-centre from, the true source position.
This bias is minimised through point-source filtering and is 
sub-dominant to the flux boosting described previously.
It is significant only for the lowest $S/N$ detections and is largely
avoided by restricting our analysis to the most robust 
sources ($S/N \gsim 4$).
The remaining small bias ($b_{peak}<$~0.2$\sigma_m$ for $S/N\geq 4$)
is estimated through simulation and subtracted from the detected
source flux ($S_m$) before calculating the PFD in Equation~\ref{eq:bayes}.

We validate these flux corrections through
extensive simulation of the PFD.
We generate 10,000 simulated maps 
by adding noiseless sky realisations to random noise maps 
using the prescription outlined in Paper I.  
Sources are randomly injected spatially (i.e. no clustering)
and in accordance with the number counts prior assumed.
We group recovered sources in the resulting maps according 
to their measured values ($S_m, \sigma_m$), with each being 
mapped back to an intrinsic flux, $S_i$, 
defined as the maximum input flux found
within $\sigma_{beam}= 7.6''$ of the output source location.  
For each bin of measured values ($S_m, \sigma_m$), 
the input $S_i$ values are binned and normalised to produce
a simulated PFD.   

Example simulation results are presented in Figure~\ref{fig:pfd1}.
Overall, the Bayesian approximation of the PFD (solid curve) provides a good 
estimate of the simulated probability distribution (histogram) at most fluxes. 
The differences at low flux and low $S/N$ 
are due to a combination of source confusion in the simulations, 
higher-order effects of the bias to peak locations, and other 
low level systematics.
For the purposes of this paper, the Bayesian results are 
preferred to the simulated PFDs for their 
computational speed, resolution, and flexibility in priors.    
The strong differences between the simulated probability distributions (histograms)
and the naive Gaussian distributions (dashed curves) 
demonstrate the significance of flux boosting in surveys of this type.   
It is important to note that flux boosting (as described above) is not 
related to the adopted detection threshold
and that even the most robust detections can be
significantly biased.  
For example, a source detected at $S/N = 8$ in the AzTEC/COSMOS map
will have been boosted by an average of 1.2~mJy ($\sim1\sigma_m$), 
assuming the scaled SCUBA/SHADES SMG population.   

\subsection{\label{ssec:countssub}Number Counts Derivation}

The relative robustness of each source candidate  
is encoded in the PFD
and is a function of both $S_m$ and $\sigma_m$,
as opposed to merely $S_m$/$\sigma_m$,
due to the population's steep luminosity function.  
As in Paper I, we use the total probability of a source candidate
being de-boosted to negative flux as the 
metric of relative source robustness.  
\citet{coppin06} found that
$P(S_i\le 0|S_m,\sigma_m) < 0.05$ provided a natural threshold from 
which to select a large sample
of robust SMGs without including a significant number of noise peaks, 
or ``false detections''.   
This threshold also marks the point
where the Bayesian approximation begins to suffer from low-level 
systematics, as suggested by the comparison of the 
Bayesian and simulated PFDs (Figure~\ref{fig:pfd1}).  
Therefore, we will use this ``null threshold'' of 5\% 
to define our catalogue of robust sources from which to estimate number counts.  
This threshold is equivalent to $S/N$ values of 4.1~-~4.3   
for our range of $\sigma_m$ values, 1.2~-~1.4~mJy, assuming
the scaled SCUBA/SHADES prior.

We derive the number counts from the catalogue of robust sources and
their associated PFDs using a bootstrap sampling 
method similar 
to that used in \citet{coppin06}. 
In each step of this 
method, the selected sources are 
randomly assigned fluxes according to their respective PFDs 
(Equation \ref{eq:bayes}).  
These samples are binned by flux
to produce differential ($dN/dS$) and integral ($N$($>S$)) 
source counts, with each bin being appropriately scaled 
for survey completeness and area.  
We introduce sample variance by sampling the robust source catalogue
\textit{with replacement} \citep[e.g.][]{press92}, and by Poisson deviating the number of 
times the catalogue is sampled around the true number of detections.
We repeat this process 20,000 times to determine uncertainty and
correlation estimates for the number count bins.

Applying this sampling method to relatively small source catalogues results in 
a discretely sampled probability distribution for each number counts bin.  
This finite multinomial distribution can be non-Gaussian and asymmetric; 
therefore, we describe the 
uncertainty in the number counts 
as 68\% confidence intervals that are approximated by linearly
interpolating between the occupation numbers sampled in the bootstrap.  

Survey completeness is estimated through simulation, in which
sources of known intrinsic flux are randomly injected into noise map realisations one 
at a time and
their output is tested against the null threshold source definition.  
Independent simulations confirm that this method provides excellent completeness
estimations at all
fluxes considered and that source confusion is not an issue given our
beamsize (18'') and map depth ($\sigma >$~1.1~mJy).
Completeness is calculated as a function of intrinsic flux and 
averaged across the map to account for the slightly varying 
depth across the survey region considered.  
We calculate the effective completeness of each differential number counts bin
by averaging the simulated completeness function within the bin,
weighted by the assumed relative abundance of sources (i.e. the prior).

\begin{figure}
\epsfig{file=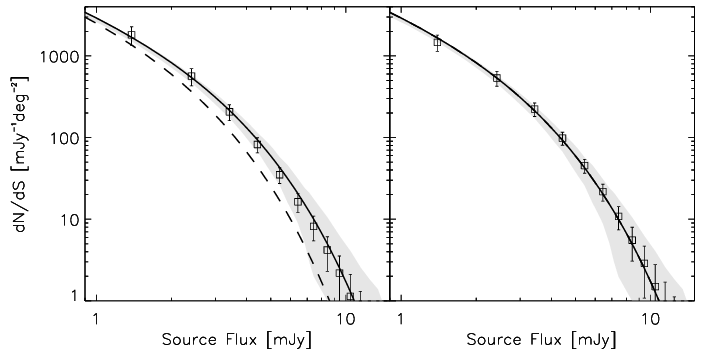, width=1.0\hsize,angle=0}
\caption{
  Simulated differential number count results (data points) using the 
  extraction techniques described in \S~\ref{ssec:countssub}.
  The solid line represents the analytic source counts (Equation~\ref{eq:schechter})
  used to populate the simulated maps, while the shaded region is the 1$\sigma$
  dispersion incurred by randomly populating maps of this size (0.15 deg$^2$).  
  The left panel shows the results when assuming a prior based on the scaled 
  SCUBA/SHADES results (dashed curve), while the right panel
  is for the ``ideal'' prior that matches the underlying input population (solid curve).
  Error bars represent the 
  1$\sigma$ dispersion from 1,000 simulations, while the errors in the means  
  are typically smaller than the data symbols plotted.
  }
\label{fig:simnc1}
\end{figure}
  
We test these techniques by applying the same flux correction and number counts extraction 
algorithms to simulated maps with the same size and 
noise properties as those of the AzTEC/COSMOS survey.  
Figure~\ref{fig:simnc1} shows the extracted differential number counts  
from simulated maps using two different assumed priors.  
Both sets of simulated maps were populated 
with the same SMG population (solid line), which is similar to the 
final results of this AzTEC/COSMOS survey (\S~\ref{ssec:countscos}).  
The right panel of Figure~\ref{fig:simnc1} shows the results of the 
ideal case where the Bayesian prior is the same distribution used to 
randomly populate the simulated maps, while the left panel shows the results
when using the scaled SCUBA/SHADES prior (dashed curve), which 
differs from the simulated input population (solid curve).
For both priors  
the extracted number counts 
are in excellent agreement with the injected population.
The relatively small differences between the input and output counts in the ideal case
are used as systematic correction factors 
in our final calculations.
The lowest flux bin (1-2~mJy) suffers from very low (and poorly defined) completeness 
and is, in general, the most sensitive to the assumptions in the prior.  
For these reasons we will restrict our analysis
in this paper to the number count results for fluxes $> 2$~mJy, unless
otherwise specified. 

In Figure~\ref{fig:simnc1}, 
the dispersion of the output source counts (error bars) 
is noticeably smaller than the dispersion of
input source counts (shaded region) at high fluxes.  
This discrepancy reflects the correlation between
output data points through our assumed prior.  
We characterise the overall bias to the assumed population
by testing a wide range of priors 
against a static input population.
For priors that are consistent with previous SMG surveys 
\citep[e.g.][]{laurent05,coppin06,scott06}, 
the bias incurred is generally smaller than the formal 1$\sigma$ errors of the 
extracted counts in a survey of this size and depth.  
Larger biases can result for exceptionally poor priors
(e.g. $>$~order of magnitude differences from the true population); 
however, in most cases the extracted number counts better 
represent the actual source population
than the initial prior,
making it possible to mitigate this bias through 
an iterative process that adjusts the prior
based on the extracted counts.
We apply this iterative method to the AzTEC/COSMOS number 
counts estimate in the next section.

\subsection{\label{ssec:countscos}AzTEC/COSMOS Number Counts}

\begin{figure}
\epsfig{file=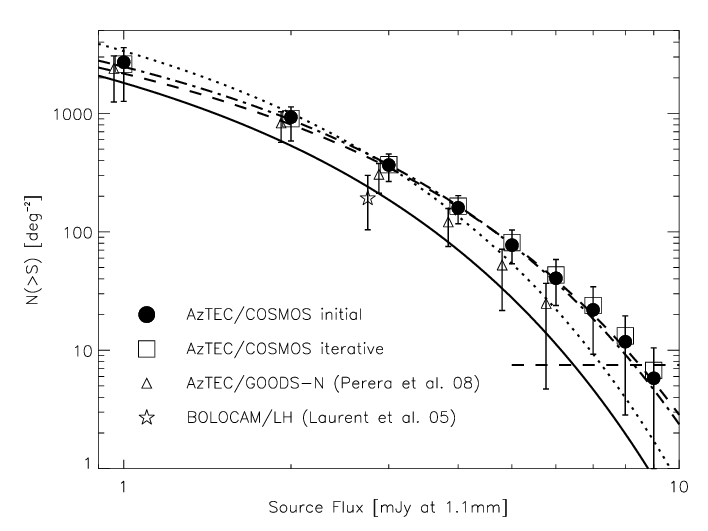, width=1.0\hsize,angle=0}
\caption{AzTEC/COSMOS integral source counts
  derived from the most robust AzTEC sources in the field using the
  techniques described in \S~\ref{ssec:countssub}. 
  Filled circles (confidence bars omitted) 
  represent the extracted counts using the initial 
  assumed scaled SCUBA/SHADES prior (solid curve).
  Empty squares and 68\% confidence bars represent the extracted counts 
  after iteratively adjusting the prior
  to best represent the results of this survey. 
  The dashed, dash-dotted, and dotted curves represent differential 
  number counts fits to Eq.~\ref{eq:schechter}
  with free parameters [$S^\prime$,$N^\ast$], [$S^\prime$], and [$N^\ast$], 
  respectively.  
  The horizontal dashed line represents the ``survey limit'', defined
  here as the source density that will Poisson deviate to zero sources 
  (per 0.15~deg$^2$) 32.7\% of the time.
  Open triangles and associated error bars represent the number counts
  found in the AzTEC GOODS-N survey \citep{perera08} using the same 
  techniques and have been shifted 4\% to the left for clarity. 
}
\label{fig:intnc1}
\end{figure}

The AzTEC/COSMOS integrated number counts are shown in Figure~\ref{fig:intnc1}.  
This field shows an excess of sources at all fluxes  
when compared to the scaled SCUBA/SHADES results (solid line).  
The number counts results are relatively insensitive to the choice of prior, 
with the initial analysis (filled circles) in agreement with those produced
using an iterative prior (open squares).  
The ``robust'' source criterion of $P(S_i<0|S_m,\sigma_m) < 0.05$ is somewhat more sensitive
to the chosen prior, with the equivalent $S/N$ threshold in a $\sigma_m = 1.3$~mJy
region being 4.2 and 4.0 for the initial 
and final iterative  
priors, respectively.  
The final iterative prior deems a larger number
of sources as robust compared to the initial prior due to the 
number of sources lying in this $S/N$ range. 
The corresponding effect on the survey completeness keeps this from being a 
runaway process, with the iterative number counts quickly converging within a few
iterations.  

The differential and integrated number counts of the AzTEC/COSMOS field are 
presented in Table~\ref{tab:counts}.  
We fit the differential number counts to Equation~\ref{eq:schechter} using 
Levenberg-Marquardt minimisation, 
incorporating the data covariance matrix to account
for correlations between flux bins.  
Various fits to the data are presented in
Table~\ref{tab:fits} and are shown in Figure~\ref{fig:intnc1}.  
Given the size and depth of this survey, we constrain 
the parametric fits to flux bins between 2 and 10~mJy 
to avoid bins that are poorly sampled and prone to systematic errors.  
This range of flux values is relatively insensitive to the 
power-law parameter $\alpha$ (\Eq~\ref{eq:schechter}); 
therefore, we fit the data while holding $\alpha$ constant at values of 
$-2$ and $-1$, which represent the SCUBA/SHADES result and a 
pure exponential, respectively.
We also present similar fits to the SCUBA/SHADES number counts \citep{coppin06}
for comparison.
The parametrised AzTEC/COSMOS results 
provide the maximum constraint on differential source counts
at fluxes $\sim$~4-5~mJy (depending on the parametrisation).  
For example, a two parameter ($S^\prime$,$N^\ast$; $\alpha$ fixed to $-2$) 
fit to \Eq~\ref{eq:schechter} constrains the AzTEC/COSMOS differential counts
at 4.5~mJy to $84\pm17$~deg$^{-2}$~mJy$^{-1}$.    

Uncertainty in the flux calibration of the AzTEC/COSMOS survey is not included
in these calculations and we believe it to be sub-dominant to the formal errors
of the source flux, number counts, and fitted parameters.  
Calibration error estimates for individual observations 
during this observing season are 6-13\% \citep{wilson08a}.
Any normally distributed random component of
this error will be reduced in the final co-added map since  
this survey is composed of multiple observations spanning many 
nights/weeks and calibration measurements.
Systematic error in the calibration is believed to be dominated by 
the 5\% uncertainty in the flux density of our primary calibrator, 
Uranus \citep{griffin93}.

\begin{table}
\begin{center}
\begin{tabular}{cccc}
\hline
Flux Density & dN/dS                   & Flux Density       &  N($>$S)    \\
(mJy)        & (mJy$^{-1}$deg$^{-2}$)    & (mJy)              &  
(deg$^{-2}$)\\
\hline
1.40 & $1706.^{+666.}_{-1534.}$ & 1.0 & $2610.^{+987.}_{-1346.}$ \\
2.42 & $535.^{+235.}_{-292.}$ & 2.0 & $904.^{+232.}_{-319.}$ \\
3.43 & $204.^{+88.}_{-82.}$ & 3.0 & $369.^{+85.}_{-103.}$ \\
4.43 & $84.^{+37.}_{-33.}$ & 4.0 & $165.^{+37.}_{-48.}$ \\
5.44 & $38.^{+18.}_{-18.}$ & 5.0 & $81.^{+22.}_{-27.}$ \\
6.44 & $19.^{+10.}_{-12.}$ & 6.0 & $43.^{+15.}_{-19.}$ \\
7.44 & $11.^{+6.}_{-9.}$ & 7.0 & $24.^{+11.}_{-15.}$ \\
8.44 & $6.5^{+4.2}_{-6.0}$ & 8.0 & $13.^{+6.}_{-10.}$ \\
9.44 & $3.9^{+3.8}_{-3.6}$ & 9.0 & $6.7^{+3.7}_{-6.7}$ \\
\hline
\end{tabular}
\end{center}
\caption{\label{tab:counts} AzTEC/COSMOS 
differential and integral number 
counts using iterative adjustment of the prior.
The differential number counts flux bins are 1~mJy wide
and span interger flux values,  
with effective bin centres (first column) weighted 
according to the assumed prior.  
The lowest flux bin listed is particularly sensitive to
uncertainties in the prior and such systematics are not 
included in the given 68\% confidence intervals.
}
\end{table}

\begin{table}
\begin{center}
\begin{tabular}{lcccc}
\hline
Data Set & $S^\prime$    & $N^\ast$      & $\alpha$ & $\chi^2$ \\
         & (mJy)       & (deg$^{-2}$)  &	    &          \\
\hline
Az/COS	 & $1.83\pm0.41$ & $4420\pm2720$ & $-2$     & 0.21     \\ 
Az/COS	 & $1.72\pm0.12$ & $5200$        & $-2$     & 0.28     \\
Az/COS	 & $1.36$        & $9610\pm1970$ & $-2$     & 1.89     \\
SHADES	 & $3.36\pm0.49$ & $5200\pm1790$ & $-2$     & 0.23     \\
\hline
Az/COS	 & $1.31\pm0.23$ & $3570\pm1790$ & $-1$     & 0.59     \\
SHADES	 & $2.39\pm0.27$ & $4370\pm1170$ & $-1$     & 0.21     \\
\hline
MODEL$_{1.1mm}$    & $1.34$        & $5280$        & $-2$     & $--$     \\
\hline
\end{tabular}
\end{center}
\caption{\label{tab:fits} 
Fit results to the differential number counts 
and respective covariance matrix
of the AzTEC/COSMOS ($1100~\micron$) and SCUBA/SHADES
($850~\micron$; \citealp{coppin06}) surveys.  
All fits are to the modified Schechter function given in Equation~\ref{eq:schechter}
while holding various parameters  
constant (those with no uncertainty given).
To avoid strong systematics at low flux, 
all fits are limited to data with $S_{1100\micron} \ge 2$~mJy and 
$S_{850\micron} \ge 4$~mJy for the AzTEC and SCUBA surveys, 
respectively.   
The last row represents our assumed 1.1~mm blank-field model
for the initial prior (scaled SCUBA/SHADES).  
$\chi^2$ values are unrealistically low, likely due
to a combination of: (a) our uncertainties being assumed as Gaussian in the fit;
and (b) additional correlation not accounted for in the linear Pearson covariance 
matrix constructed through the bootstrap sampling method (\S~\ref{ssec:countssub}).  
These effects are also seen in the SCUBA/SHADES implementation
of this algorithm \citep{coppin06}.  
}
\end{table}

\subsection{\label{ssec:blankfield}Blank-Field Model}

It is immediately apparent that the AzTEC/COSMOS field
is rich in bright sources 
when compared to other 1.1~mm surveys (see \S~\ref{ssec:countsod}).
In order to quantify the significance
of this potential over-density, we must first
adopt an accurate characterisation of  
the true 
background (``blank-field'') population.
The tightest published constraint on the SMG
population is provided by the 
850~$\micron$ SCUBA/SHADES survey \citep{coppin06}, 
which we convert to 1100~$\micron$ assuming an
$850~\micron$/$1100~\micron$ power-law spectral index of $3.5$.  
This scaling is roughly consistent with the 
integrated number counts of the 1.1~mm 
Bolocam Lockman Hole survey \citep{laurent05}, 
which partially overlaps with the SCUBA/SHADES
survey.
Assuming the SCUBA and AzTEC observations 
are in the Rayleigh-Jeans regime of 
optically-thin thermal dust emission from the SMGs,
our scaling is consistent with the sub-mm spectral indexes
of bright IR galaxies in the local universe 
\citep{dunne00,dunne01}.    

Using a scaled version of the number counts measured 
at a different observation wavelength carries the
inherent risk that the two bands are sensitive to significantly 
different (although overlapping)
source populations, as evidenced by the possible existence of 
sub-millimetre drop-outs \citep[SDOs, ][]{greve08}.  
The SCUBA 850~$\micron$ surveys would be relatively insensitive to 
these proposed SDOs due to
a combination of high redshift ($z \gg 3$) and/or unusual spectral 
energy distributions (e.g. $T_{dust} \sim 10$~K).
Therefore, it is important to verify the blank-field model
with a direct measurement of 1.1~mm population.

The most robust characterisation of the AzTEC/COSMOS over-density
comes through comparison to similar analyses of other AzTEC 1.1~mm surveys, 
which eliminates systematics between different instruments and minimises 
those related to calibration. 
The best AzTEC 1.1~mm blank-field constraints are being provided 
by the AzTEC 0.5~deg$^2$ survey of the SHADES fields.  
Initial results of the AzTEC/SHADES survey
(using nearly identical algorithms as those applied to AzTEC/COSMOS)
are consistent with our scaling of the SCUBA/SHADES counts.
Our number counts model falls in the higher regions of the 
AzTEC/SHADES uncertainty interval 
(modelled differential counts are roughly $+$0.5$\sigma$ to $+$2.0$\sigma$ above
the average AzTEC/SHADES counts in the flux range explored here)
and is within the field-to-field variations measured in those 
large surveys; 
therefore, we believe our 
model represents a conservatively high estimate of the 
blank-field counts that is appropriate for   
robust qualification of the potential over-density. 

We note that the AzTEC survey of the 
GOODS-N field \citep{perera08} 
finds an SMG number density that is somewhat higher than
our blank-field model; 
however, our model is within 
the $\sim 1\sigma$ uncertainty of that survey's integrated number counts  
(Figure~\ref{fig:intnc1}; note that the data points are correlated).   
The AzTEC/GOODS-N results imply an $S^\prime$ parameter 
($S^\prime = 1.25 \pm 0.39$~mJy) that is consistent
with our general scaling of the SCUBA/SHADES counts,  
but suggest systematically higher number counts 
(i.e. larger $N^\ast$ parameter).
The small size of the GOODS-N survey
(0.068~deg$^2$) 
makes it highly susceptible to cosmic variance and clustering, 
thus reducing it's viability as a measurement of the average sky.  
It also does not significantly 
constrain the bright ($S>$~5~mJy) 1.1~mm source counts where the 
AzTEC/COSMOS over-density is most apparent (\S~\ref{ssec:countsod}).
The SCUBA survey of GOODS-N \citep{borys03} already suggests the field may 
be overly rich in sub-millimetre sources, with number counts systematically higher
than seen in the SCUBA/SHADES blank-field \citep{coppin06}, 
although the analyses of these two 
SCUBA surveys differ significantly and the difference 
in number counts could be partially systematic.

\subsection{\label{ssec:countsod}SMG Over-density}

The source catalogue presented in Paper I  
suggests that the AzTEC/COSMOS field has a significantly larger number of bright 
1.1~mm sources than might otherwise be expected for a survey of this size
and depth.
The density of sources
in the AzTEC/COSMOS field with raw measured fluxes 
$\ge6$~mJy is 3 times higher
(14 sources in 0.15 sq. deg. field) than 
in the
1.1~mm Bolocam Lockman Hole survey of similar depth
(3 sources in 0.09 sq. deg.; \citealp{laurent05}).
The seven brightest AzTEC sources in the COSMOS field have been imaged 
using the Submillimeter Array (SMA) at 890~$\micron$,
and they are shown to be single, unresolved sources at 2 arcsec
resolution \citep{younger07}.  

We compare the AzTEC/COSMOS 
number counts to the blank-field model 
discussed in \S~\ref{ssec:blankfield}.  
Figure~\ref{fig:intnc1} shows that 
the AzTEC/COSMOS 
integrated source count
estimates are clearly in excess of the scaled SCUBA/SHADES counts.  

\begin{figure}
\epsfig{file=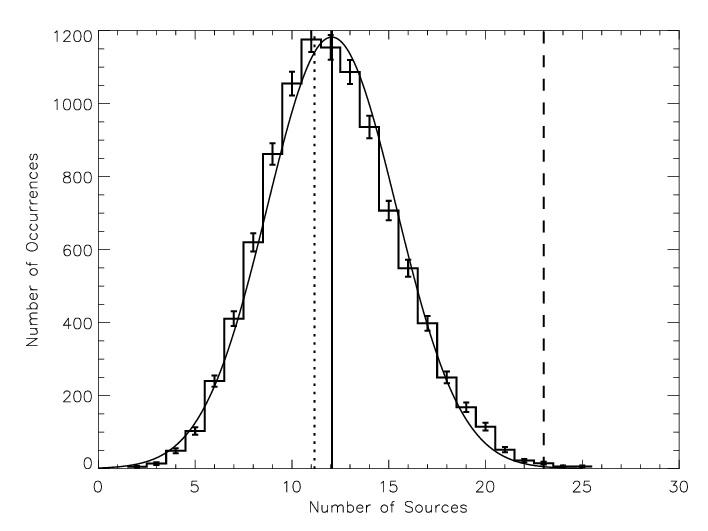, width=1.0\hsize,angle=0}
\caption{Distribution of the number of robust sources detected in 10,000 
simulations (using 300 unique noise realisations) 
of the AzTEC/COSMOS observations when randomly populating the 
astronomical sky with the scaled SCUBA/SHADES number counts 
(solid histogram).  Using the same source criteria,  
there are 23 robust sources detected in the AzTEC/COSMOS map
(dashed vertical), which is greater than 99.7\% 
of the simulations.  A Gaussian fit to the simulation results
(thin solid curve) shows 23 sources to be a 3.3$\sigma$ outlier.  
The difference between the simulation mean (solid vertical) and the
semi-analytic expectation value (dotted vertical) reflects the 
number of false detections (i.e. $\sim$~1 per map) for the scaled SCUBA/SHADES
prior and chosen source threshold.  
}
\label{fig:intod1}
\end{figure}

To estimate the probability of this excess happening by chance, we
compare the number of robust sources detected in the AzTEC/COSMOS survey 
to the number recovered in simulated maps.
In Figure~\ref{fig:intod1} we show the distribution of the number of 
recovered sources in 10,000 simulations, each populated with a random
realisation of the scaled SCUBA/SHADES counts.
On average, 12.1 sources are recovered from each of the simulated maps, 
which is in agreement with the semi-analytical expectation value of 11.2 
(calculated from the scaled SCUBA/SHADES results and 
simulated completeness of this survey)
and the expected number of false detections 
($\langle N_{false} \rangle \approx 1.2$). 
Application of the same source criteria (5\% null threshold, scaled 
SCUBA/SHADES prior) to the real map results in 23 robust sources
(32 if using the iterative prior), 
which is greater than in 99.7\% 
of the simulations.
The AzTEC/COSMOS source over-density is even more significant in 
the number of very bright sources,
with 11 detected at $S/N~\ge$~5 ($S_m~\gsim$~6.2~mJy).
Ten-thousand simulations of the blank-field model could produce
no more than 6 such detections in a single map,
thus inferring a $\gg 4\sigma$ significance 
in the number of bright sources.

Assuming the scaling of the SCUBA/SHADES counts 
accurately represents the blank-field
SMG population at 1.1~mm, the 
parametric fits shown in Table~\ref{tab:fits} favour 
the over-density being described as 
a shift in the flux parameter $S^\prime$ over an 
increase in 
the normalisation parameter $N^\ast$. 
This is consistent with an apparent over-density caused by uniform 
amplification of the source fluxes.  
However, this solution is degenerate
with an alternative scaling of the SCUBA/SHADES results -- 
scaling with a flatter spectral index of $\sim2.6$ also produces a good 
fit to the AzTEC/COSMOS number counts 
(dash-dotted curve of Figure~\ref{fig:intnc1}).
Therefore, an accurate representation of the blank-field
population (\S~\ref{ssec:blankfield}) is critical to properly quantifying the over-density.  

If taken alone, the over-density of SMGs in 
the AzTEC/COSMOS field would likely be explained away as 
simple cosmic variance in the SMG population 
as traced in a 0.15~deg$^2$ field. 
However, in the following sections we demonstrate 
that the over-density is due, in part, to 
\textit{foreground} structure in the COSMOS field.  
Only with the rich multi-wavelength coverage 
of the COSMOS field and the relatively large size of
the AzTEC map is this analysis possible.

\section{Correlation Between AzTEC sources and Large Scale Structure 
in the COSMOS field}
\label{sec:LSScorr}

Having shown that the AzTEC/COSMOS field exhibits a
significant excess of bright SMGs with respect to 
our adopted blank-field model, 
we explore the possibility that this over-density is due, in part, to
a correlation of AzTEC sources with the prominent large-scale
structures at $z\lsim 1.1$ identified in this portion 
of the COSMOS field \citep{scoville07b}.
All correlation tests in this section are limited to the inner
0.15~deg$^2$ region of the AzTEC map where the uniformity in 
coverage simplifies the analysis.

\begin{figure}
\epsfig{file=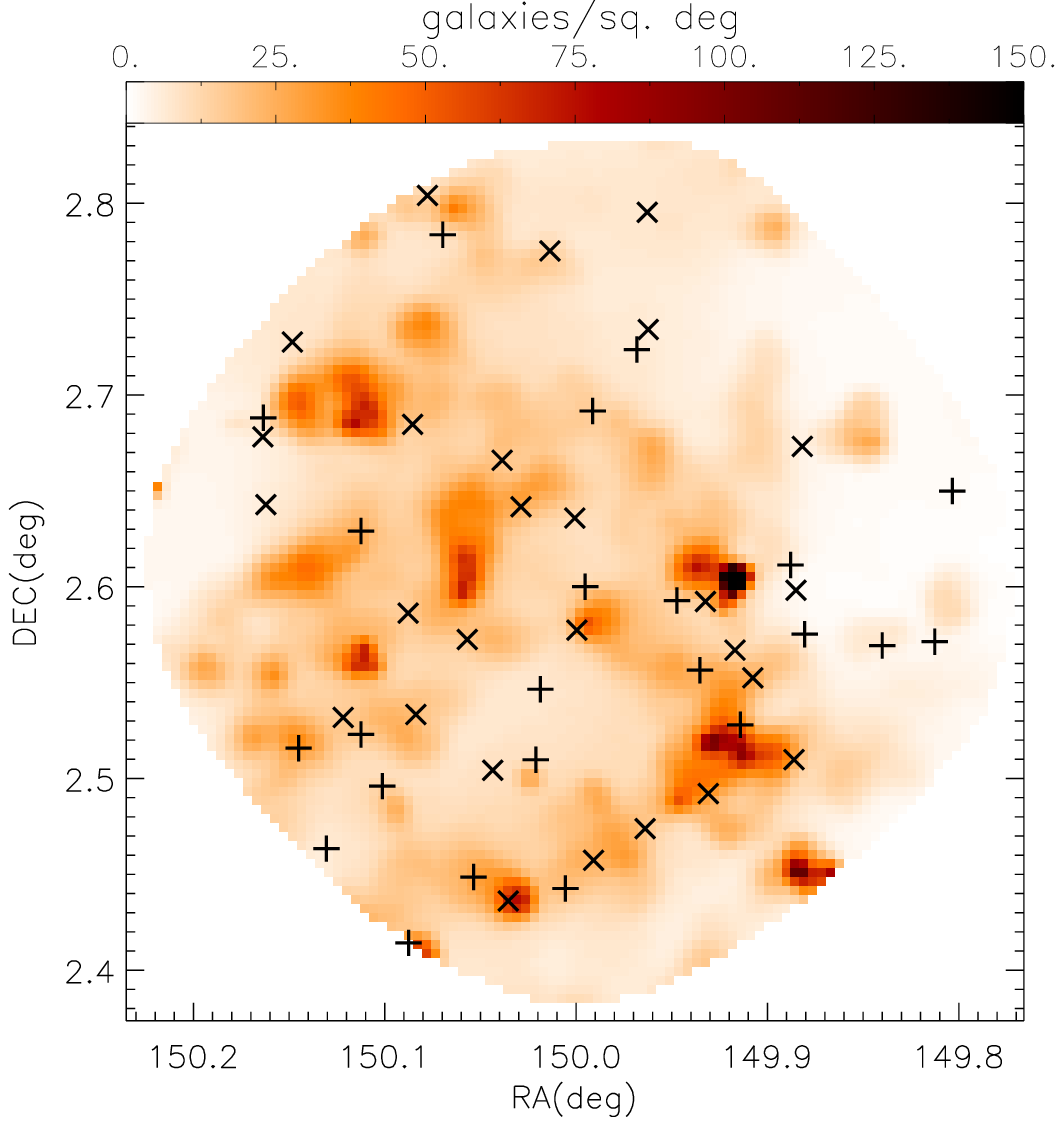,width=1.0\hsize,angle=90} 
\caption{Smoothed surface density map of galaxies derived from the
  optical-IR catalogue of COSMOS galaxies \citep{scoville07b} in the
  0.15 sq. deg area surveyed by AzTEC, where darker colours indicate
  more densely populated areas of the sky. The cross and plus symbols
  represent AzTEC sources detected at signal-to-noise ratios $S/N\ge4$
  and $4>S/N\ge3.5$  \citep{scott08}, respectively.  
  }
\label{fig:galden1}
\end{figure}

The smoothed galaxy density map produced by the COSMOS consortium
(\citealp{scoville07b}, see Figure~\ref{fig:galden1}) shows a
collection of dense regions in the 
the AzTEC covered area. 
We first look for coincidence with AzTEC sources 
by cross-correlating the surface-density of optical-IR galaxies in this map 
with the AzTEC source positions using a bi-dimensional Kolmogorov-Smirnov (K-S)
test of similarity \citep{peacock83,fasano87}.
Restricting this analysis to the 30 
robust AzTEC sources detected at a $S/N\ge 4$, which
have an estimated false detection rate of less than 7\% \citep{scott08},
we find that the test cannot reject the null hypothesis
that the distribution of AzTEC-source positions follows the
surface density distribution of optical-IR galaxies.  
The test concludes that the difference between the SMG and 
optical-IR populations 
is smaller than 93.7\% of the differences expected at random
due to sample variance, 
often referred to as rejecting the null hypothesis at the 6.3\% level, 
thus suggesting the distributions could indeed be similar.  

The significance of the SMG positional correlation with the large-scale structure 
is further quantified by comparing the K-S $D$-statistic of the 
SMG catalogue to that of a homogeneous random distribution of the same number of sources,
under the null hypothesis
that they follow the surface density of optical-IR galaxies. 
This test determines that the AzTEC/COSMOS source distribution 
follows the optical-IR distribution more strongly than 98.9\% 
($\sim$~2.5$\sigma$)
of the random-position catalogues.  
The result is somewhat less significant, 91.1\%,
if we expand the comparison to the full $S/N\geq 3.5$ AzTEC/COSMOS
catalogue, which is likely due to the increased number of false detections 
(from $\sim$~1 to $\sim$~11)
at this lower $S/N$ threshold.  
These false detections (noise peaks) are inherently random and homogeneous 
in their distribution and dilute the correlation signal.

\begin{figure}
\hspace*{-0.7cm}
\epsfig{file=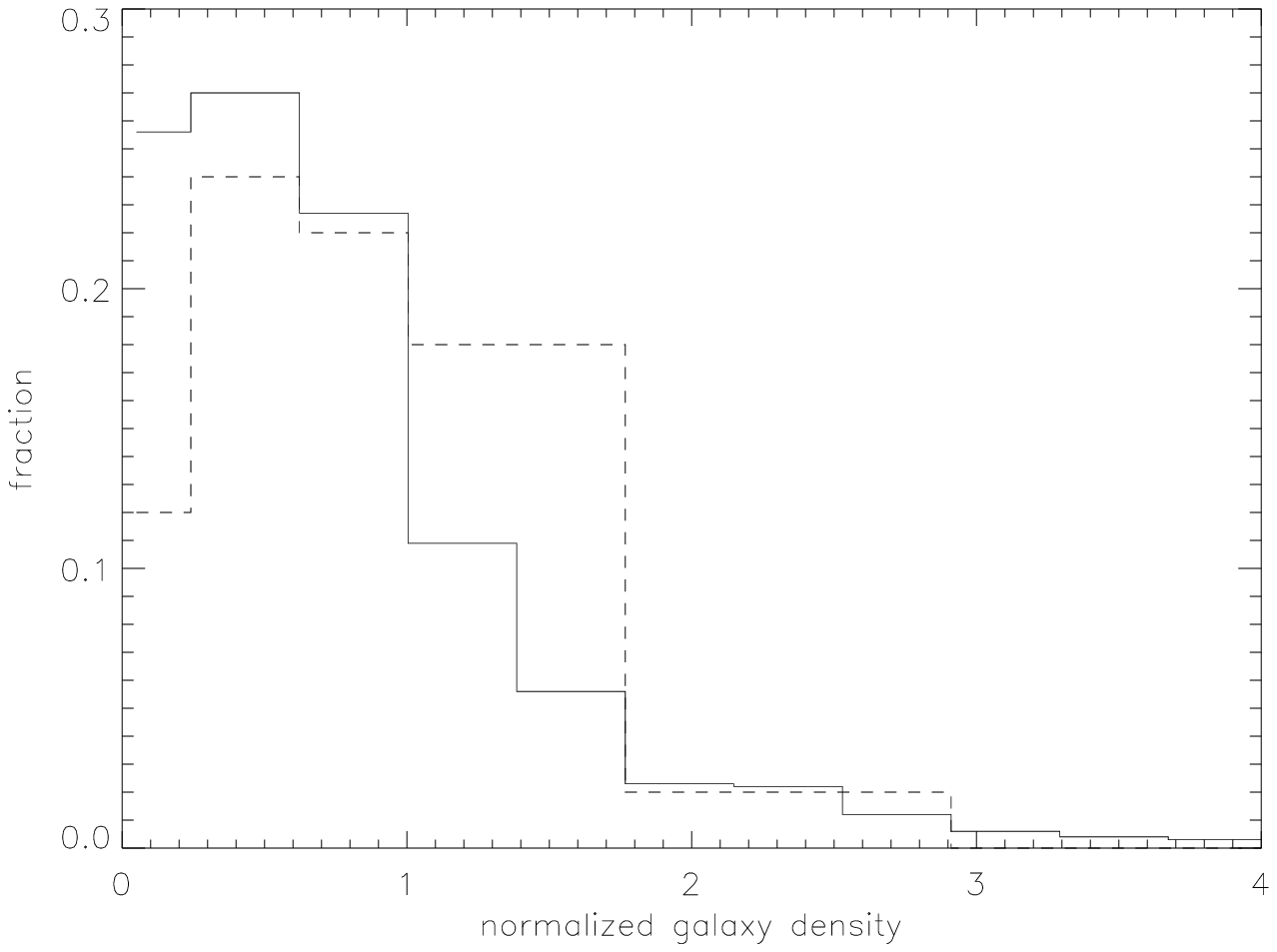,width=0.8\hsize,angle=90} 
\caption{Histogram of the galaxy density at
  $z\lsim1.1$ found within 30~arcsec of (a) AzTEC/COSMOS source candidates
  detected at $S/N\ge 3.5$ (dashed line) and (b) random positions
  in the AzTEC mapped area of COSMOS (solid line). 
  The two populations are different at the 97.2\% confidence level, 
  using a one-dimensional K-S test.
  The galaxy densities are normalised to the mean galaxy density in the full
  AzTEC-covered area.}
\label{fig:histo_galden1}
\end{figure}

It is possible that only a fraction of the AzTEC source positions are correlated
with the prominent large-scale structures detected in the COSMOS galaxy
density map while a subset of randomly distributed source positions
dilutes the sensitivity of the quadrant-based bi-dimensional K-S statistic 
discussed above.  
Therefore, we further test the hypothesised correlation by comparing the
surface density of optical-IR galaxies within a small area surrounding AzTEC
positions to that surrounding random positions in the map.
Figure~\ref{fig:histo_galden1} shows the distribution of the galaxy densities
at $z\lsim1.1$ projected within 30'' (1.7 pixels in
the smooth galaxy density map) of the AzTEC
source positions, compared to the galaxy densities found around 
random positions within the AzTEC survey region.
The two distributions are clearly different, with a
one-dimensional K-S test rejecting the null hypothesis of identity at 
$>$~99.99\% and 97.2\% levels for the $S/N\geq 4$ and $S/N\geq3.5$ catalogues, respectively. 
The mean number of nearby optical-IR galaxies at 
$S/N\geq 4$ ($S/N\geq3.5$) AzTEC source positions
is larger than that at random positions in the map at a significance of
99.99\% (99.5\%) according to the non-parametric Mann-Whitney (MW)
$U$-test.

\begin{figure}
\hspace*{-0.7cm}
\epsfig{file=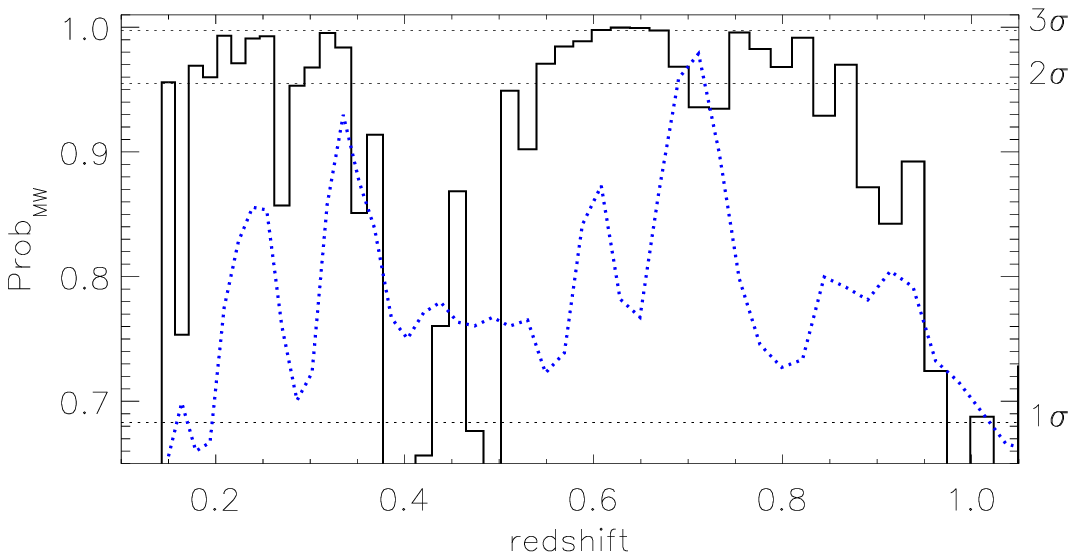,width=0.55\hsize,angle=90} 
\caption{Bar-representation of the Mann-Whitney probability that the
  mean galaxy density around 
  AzTEC sources at a given redshift slice is significantly larger
  than the mean galaxy density around random positions.   
  Horizontal dotted lines represent the $1\sigma$, $2\sigma$, and $3\sigma$
  significance levels, respectively.
  The blue-dotted curve shows the relative 
  number of optical/IR galaxies contained within each
  redshift slice within the AzTEC 
  covered area ($N_{\rm gal}(z)/N_{\rm total}*10.+0.60$).   
  }
\label{fig:z_contrib}
\end{figure}

We can search in redshift space for the structures that contribute
the most to the coincidence between AzTEC sources and the galaxy 
density in their ``line of sight'' using the photometric redshifts
of the optical-IR population \citep{ilbert08}, which have a
mean accuracy of $|\Delta z| /(1 + z) \approx$~0.01-0.02.
Figure~\ref{fig:z_contrib} shows a bar-representation of the MW probabilities
that the mean integrated galaxy density around AzTEC sources
is significantly larger than that around random positions in the map for 
various redshift slices.
There is positive signal ($\gsim 2\sigma$) arising at different
redshift slices, most notably at $z \sim 0.65$.
At redshifts $z>1.1$, the number of galaxies detected at optical-IR
wavelengths decreases significantly, and the level of correlation
found with AzTEC sources is well below the 2$\sigma$ threshold.

\begin{figure}
\epsfig{file=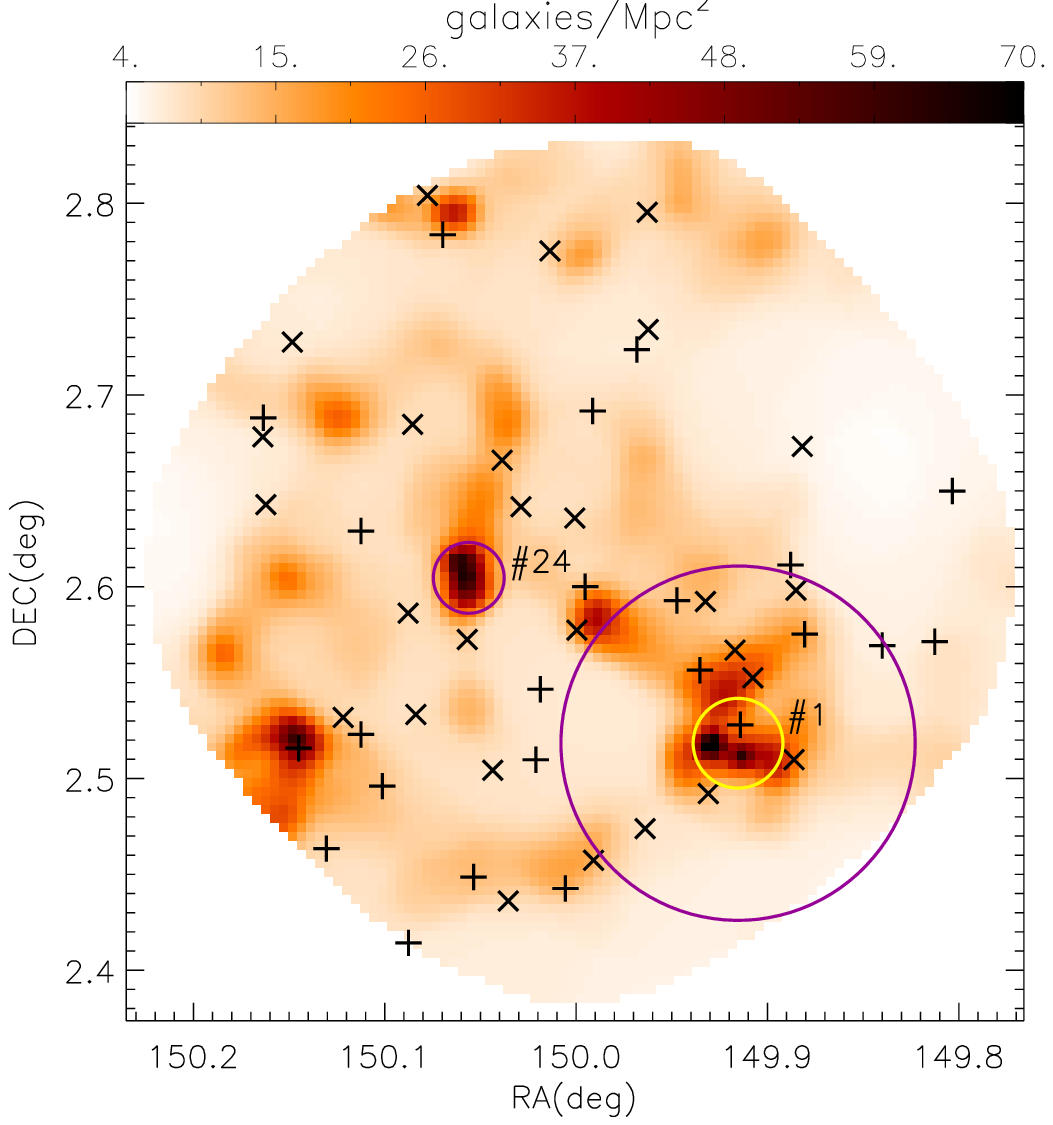,width=1.0\hsize,angle=90} 
\caption{Smoothed surface density map of galaxies at $0.60 \lsim z
  \lsim0.67$ detected at optical-IR wavelengths by the COSMOS survey
  \citep{scoville07b}.  The large-scale structure at $z=
  0.73\pm0.26$ detected by \citet{scoville07b} is marked as
  Structure \#1 and the large circle (6' diameter). 
  This large-scale structure has a peak over-density
  at $z\sim0.73$, outside of the redshift range of this figure, 
  and is identified as a massive cluster \citep{guzzo07}. The yellow circle (1.5' diameter)
  marks the spatial extent of this cluster as traced by the X-ray
  contours. Another rich cluster, at $z\sim 0.61$, is marked as Structure
  \#24.  
  Symbols are the same as in Figure~\ref{fig:galden1}.
  }
\label{fig:galden2}
\end{figure}

The most prominent contribution to the AzTEC-optical/IR correlation lies at
$0.6\lsim z \lsim 0.67$, with the redshift slices within this range 
having MW probabilities of difference 
up to 99.98\%.
The smoothed galaxy density map for this redshift range is shown in
Figure~\ref{fig:galden2}.
Two prominent
large-scale structures have been identified 
\citep[Structures \#1 and \#24 in][]{scoville07b} within this redshift slice. 
Structure \#1 at $z = 0.73 \pm 0.27$ has $1767$ optical-IR galaxy members
and approximately spans (FWHM) $\Delta$~RA~$= 0.22$~deg and
$\Delta$~Dec~$= 0.17$~deg.
Structure \#24, a less massive but very compact system that is 
X-ray detected, has 85 galaxy members and is at z~$\sim$~0.61.
Structure \#24, however, does not appear to contribute to the correlation, 
as no AzTEC sources fall within its primary extension.
Structure \#1 contains a rich core and represents a
massive cluster ($\sim 10^{15} M_\odot$) at $z\approx 0.73$,
which is clearly seen in the COSMOS weak-lensing convergence map 
\citep{massey07} and in X-ray emission
\citep{guzzo07}. This cluster 
lies outside the redshift span of strong correlation, 
but the filamentary structure that leads to it is part of the redshift 
slice under analysis (see Figure~\ref{fig:galden2}).

\begin{figure}
\hspace*{-0.7cm}
\epsfig{file=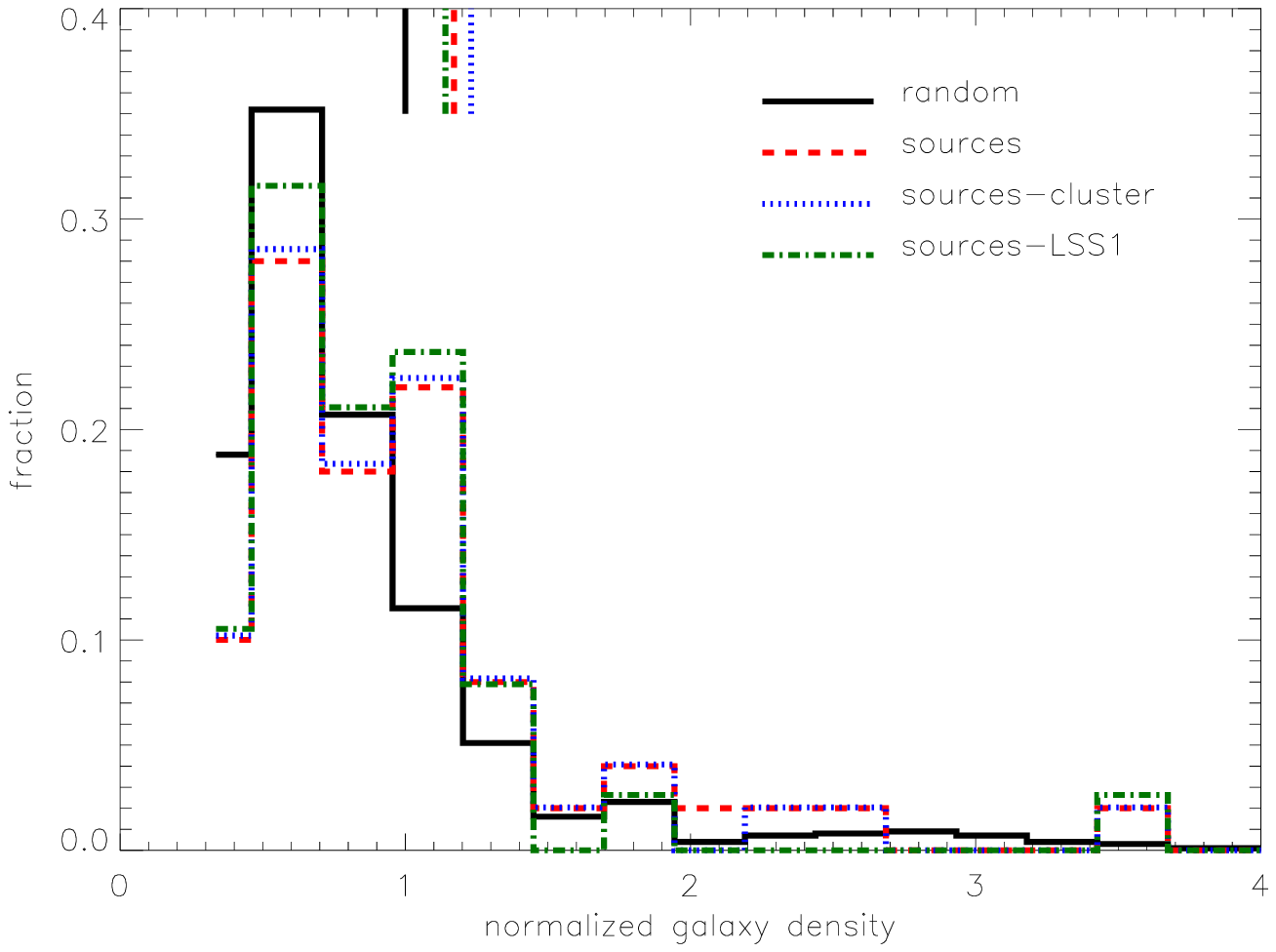,width=0.8\hsize,angle=90} 
\caption{Histogram of the fraction of optical-IR selected galaxies at
  $0.60\lsim z \lsim 0.67$ found within 30~arcsec of AzTEC source candidates
  (red dashed line), and around 
  random positions within the AzTEC-mapped
  area of COSMOS (black solid line). The blue dotted-line histogram
  represents the number of optical-IR galaxies around AzTEC sources,
  excluding the single AzTEC source that falls within the X-ray traced
  cluster-outskirt region ($\theta \lsim 1.5$~arcmin from the cluster 
  centre, \citealp{guzzo07}), 
  while the green dash-dotted histogram excludes the full 
  6~arcmin radial structure identified as Structure \#1. 
  The mean values of these histograms are represented at the top of the 
  figure as vertical bars.
  The galaxy densities are normalised to the mean galaxy density in the full 
  AzTEC-covered area such that the mean density of random positions is 1.
}
\label{fig:histo_galden2}
\end{figure}

We next assess whether the substructures within 
Structure \#1 are the main contributors to the observed correlation. 
Figure~\ref{fig:histo_galden2} 
shows the distribution of
galaxy densities around AzTEC sources and around random positions in the 
collapsed $0.6\lsim z \lsim 0.67$ map.
The means differ at the 99.8\% confidence level according to the MW $U$-test. 
If we exclude a circular region around the
cluster centre with radius $\theta = 1.5$'($\sim$~0.6~Mpc), which contains both
the cluster-core and the cluster-outskirt regions seen by the 
X-ray temperature profile \citep{guzzo07}, the significance of the
difference is 99.94\%.  
Excluding a larger circular region of radius
$\theta = 6$'($\sim$~2.4~Mpc), which represents the FWHM of the full Structure \#1, the
MW significance decreases to only 98.6\%. 
This demonstrates that
although AzTEC sources do correlate with the galaxy densities 
associated with the extended Structure \#1, the less prominent
large-scale structure across the rest of the map is also well-correlated
with the AzTEC positions.

\begin{figure}
\epsfig{file=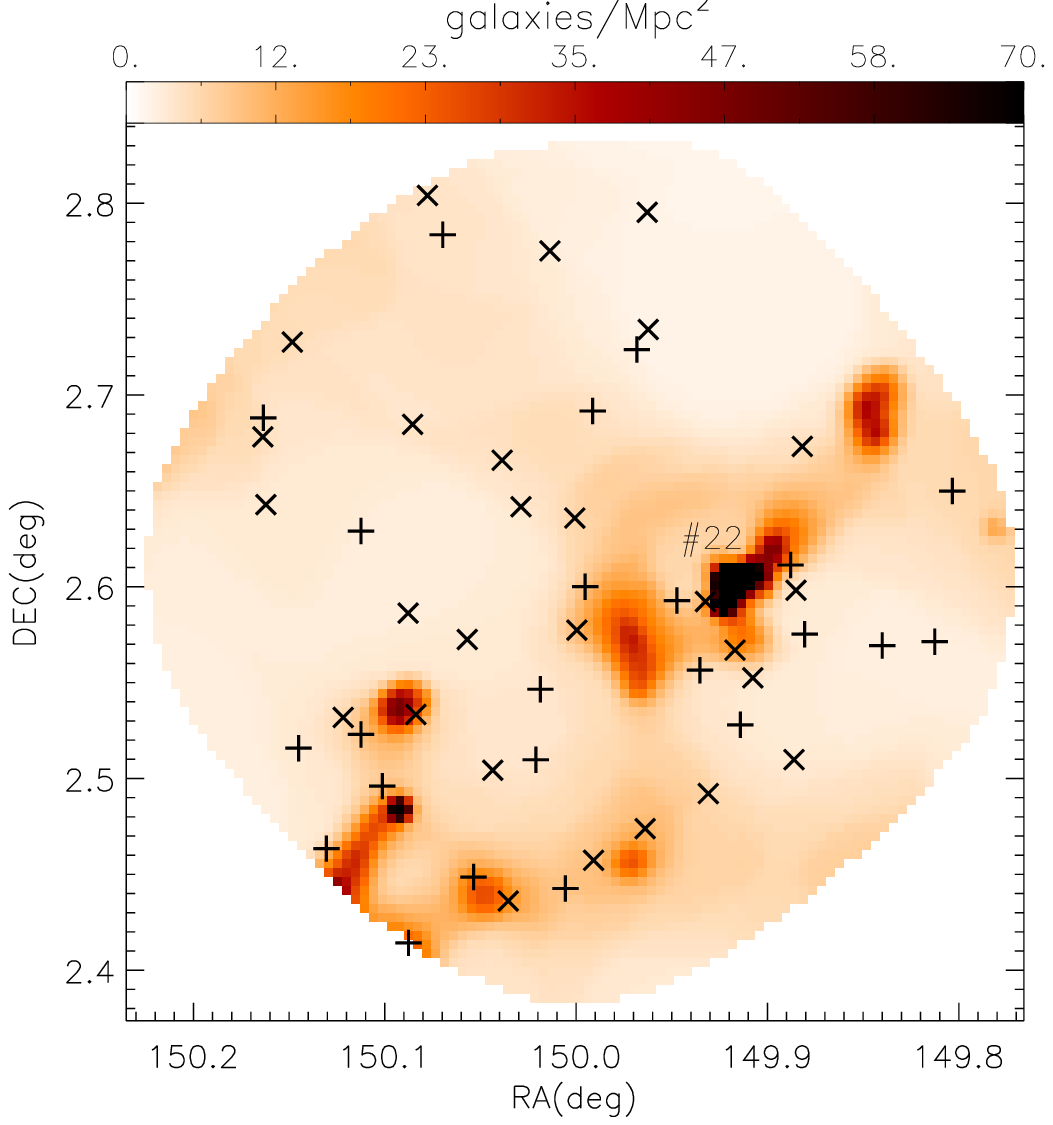,width=1.0\hsize,angle=90} 
\caption{Smoothed surface density map of galaxies at $0.24\lsim z \lsim0.26$
  detected at optical-IR wavelengths by the COSMOS survey, which 
  includes Structure \#22 of \citet{scoville07b}.  
  Symbols are the same as in Figure~\ref{fig:galden1}}.
\label{fig:galden3}
\end{figure}

Figure~\ref{fig:galden3} shows the optical/IR galaxy density map for the redshift
slice $0.24\lsim z \lsim 0.26$, which 
is also a large contributor to
the overall correlation between the large-scale-structure of the field
and AzTEC sources (Figure~\ref{fig:z_contrib}). 
Structure \#22 from \citet{scoville07b}, with
$\sim 67$ possible galaxy members at $z\approx 0.26\pm0.11$,
is the main cluster in this redshift slice and is also detected in X-ray. 
However, as with the portion of Structure \#1 in the 
$0.60 \lsim z \lsim 0.67$ slice, this
system does not dominate the overall correlation with AzTEC sources:
the mean galaxy density around AzTEC sources differs from random 
locations at the 99.1\% level after exclusion of the $\Delta$~RA~$\approx 0.06$~deg and 
$\Delta$~Dec~$\approx 0.14$~deg area of influence of the cluster.
Similarly, we find that the prominent contributions
of other redshift slices  
(e.g. $z \sim 0.33$ and $z \sim 0.8$) 
to the overall spatial correlation 
are not due to single compact structures.

It appears that the observed correlations are not dominated by the clusters
in the field, thus
it is not surprising that the AzTEC positions are, in general,
less correlated with the weak-lensing mass map of
COSMOS \citep{massey07}, which is particularly sensitive to the most massive
structures like the $z\approx 0.73$ cluster (see Figure~\ref{fig:lensmass2}).  
The null hypothesis that the distribution of masses found within 30'' 
of AzTEC positions is the same as that found around random positions 
in the weak-lensing map 
is ``rejected'' at only the 60\% level (K-S test),
and their means differ 
at the 91.5\% level (MW $U$-test).  

\section{Discussion}
\label{sec:discussion}

The distribution of AzTEC sources is correlated with the
large-scale distribution of optical/IR galaxies at $z\lsim 1.1$  
and the primary (but not unique) contributors
to this signal are located at redshifts $0.60\lsim z \lsim 0.65$ and
$0.24 \lsim z \lsim 0.26$ (section~\S~\ref{sec:LSScorr}). The correlations 
in these redshift regimes are robust, with mean optical/IR densities 
differing from random distributions 
at the 99.1 to 99.98\% level (MW $U$-test).
For the seven AzTEC sources that have
been followed up with SMA interferometry \citep{younger07} and the
additional 14 that have radio detections \citep{scott08},
secure optical/IR counterparts have been
identified. The optical-IR and FIR-mm-radio photometric redshifts of
these sources place the majority of
these objects at
$z\gsim 3$;
therefore, AzTEC detected sources are most
likely background systems to the $z\lsim 1.1$ galaxy densities
shown in Figure~\ref{fig:galden1}.
This is not surprising, 
given that the population of SCUBA SMGs with radio counterparts have 
a median redshift of 2.2 \citep{chapman03,chapman05,aretxaga03,aretxaga07,pope05}.

The amplification caused by massive clusters at intermediate redshifts
($z\sim 0.2-0.4$) has been used to detect and study the
sub-millimetre galaxy population since the first SMG surveys 
(e.g. \citealp{smail97}).
Lensing is expected to occur also in and around
the $z\approx 0.73$ cluster detected in the COSMOS field, but only 4
of the 50 $S/N\ge3.5$ AzTEC source-candidates are projected within 
2' of the dense cluster core.  
Removing these sources/regions from the  
analysis of \S~\ref{sec:LSScorr} has little effect on the 
correlation strength. 
Thus the correlations seem to be tied to the
general $z\lsim 1.1$ large-scale structure in the field.
The same result holds if we exclude 
the other prominent structures in this field, 
Structures \#24 at $z\approx 0.61$ and \#22 at $z\approx 0.26$. 
Furthermore, the bulk of AzTEC sources do not
significantly correlate with the weak-lensing map, which is particularly 
sensitive to the mass contained in rich clusters.

Lensing of the sub-mm galaxy population by foreground low-redshift
structures has been claimed in the correlation analysis of 39
sub-millimetre galaxies detected in 3 disjoint fields with the density
of $R<23$~mag galaxies \citep{almaini05}, which statistically lie
at $\langle z \rangle\sim 0.5$. 
It was argued that the bright
$S_{850\mu {\rm m}} > 10$~mJy
sources are found to cluster preferentially around the highest-density areas,
and \citet{almaini05} estimate that 20-25\% of the sub-millimetre galaxy
population is subject to lensing by foreground structures.
We note that a similar study performed in the GOODS-N region \citep{blake06}
found no detectable correlation between 35 SCUBA-selected SMGs 
and the optically-selected galaxy populations 
at $z \le 0.8$.
This difference in correlation strength may be related to cosmic
variance of foreground structure on the scale 
of these maps.   
There also exist potential cases of lensing by individual galaxies, with some
SCUBA sources being incorrectly identified
as low-redshift galaxies due to
intervening foreground galaxies 
that lie directly along the line of sight
\citep{chapman02b,dunlop04}.
Since it includes a high-density region within the COSMOS field, the AzTEC
survey is sensitive to 
all of these types of amplification, and we have
demonstrated that there is a positive correlation with the large-scale
structure.
Inspection of the optical/IR counterparts of the 21
AzTEC galaxies with radio and/or sub-mm interferometric positional accuracy,
including the 7 sources known to have sub-millimetre emission on scales
$\theta < 1.2$'' \citep{younger07},
show no obvious signs of strong galaxy-galaxy lensing  
and hence any amplification of this sub-sample
must be attributed to weak-lensing.  

If our blank-field number counts model 
(\S~\ref{ssec:blankfield}) accurately represents
the intrinsic (non-amplified) SMG population in the AzTEC/COSMOS
field, 
then the observed number density of sources  
is also consistent with weak-lensing of the background
SMG population.  
Parametric fits 
to the flux-corrected number counts 
(\S~\ref{ssec:countscos}, see also Table~\ref{tab:fits} and Figure~\ref{fig:intnc1}) 
show that the relative over-density of sources 
can be fully explained as a systematic increase in the parameter
$S^\prime$, which is consistent with an average flux
amplification (e.g. lensing) of the source population by $\sim$~30\%.
Conversely, the number counts data are only marginally 
consistent with a simple 
increase in the normalisation parameter
$N^\ast$,
thus disfavouring a uniform physical 
over-density (e.g. cosmic variance)  
of sources in this field as the sole cause of the
observed over-density.  
Additionally, any over-density due to variance and/or clustering
can not  
explain the correlation of AzTEC sources
to the $z \lsim 1.1$ structure, as the AzTEC sources are 
likely background sources and not physically associated with the
$z \lsim 1.1$ structure.

An alternative cause of the number counts over-density can be imagined as
an additive flux source (e.g. dense screen of faint foreground sources)
confused with the blank-field sources.
However, the AzTEC/COSMOS map  
has been filtered for point source detection and has a mean of zero \citep{scott08},  
which leaves the map insensitive to high-density or uniform millimetre flux
sources that span large spatial scales.
Furthermore, the positions of AzTEC sources are not strongly correlated
with the most dense and compact foreground regions (i.e. clusters)
that could otherwise be potential sources of additional mm-wave flux in our  
map (e.g. the Sunyaev-Zel'dovich effect). 

\begin{figure}
\epsfig{file=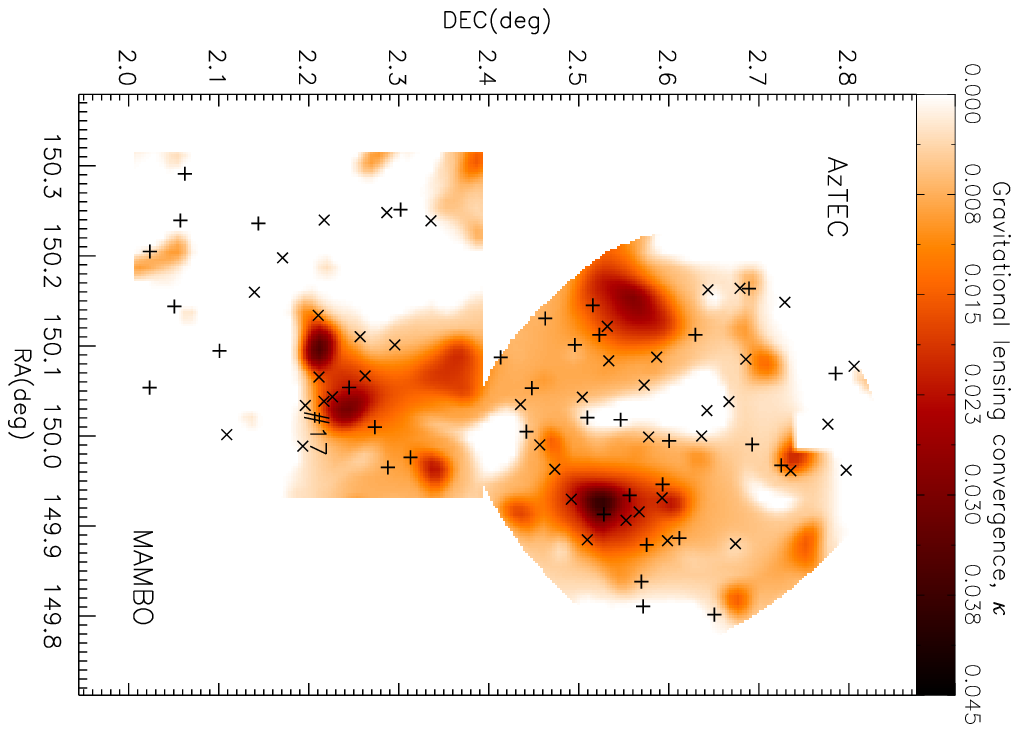,width=1.5\hsize, angle=90} 
\caption{
Weak-lensing convergence mass map \citep{massey07} of the
AzTEC and MAMBO surveyed regions of COSMOS. 
Crosses mark the millimetre source positions from the two catalogues
and follow the notation of
Figure~\ref{fig:galden1}.
}
\label{fig:lensmass2}
\end{figure}

The significance of the spatial correlation between AzTEC sources and the
intervening large-scale structure contrasts with the lack of a similar
detectable signal among the sources discovered by COSBO \citep{bertoldi07}, 
the 1.2~mm MAMBO survey to the south and adjacent to the AzTEC surveyed 
area (see Figure~\ref{fig:lensmass2}). 
If we repeat the analysis performed in \S~\ref{sec:LSScorr} with the
MAMBO catalogue, we do not find a significant correlation with the
COSMOS optical/IR galaxy surface density; the probability that the 
galaxy densities around
MAMBO sources are different from that around random positions in the map
is only 87\% (K-S test). 
This lack of a significant correlation signal may be due in part
to the smaller catalogue of significant sources in the COSBO 
field and the overall lack of significant foreground structure 
in much of the COSBO covered area.  

The association between COSBO sources and the weak-lensing
derived mass-map, however, is stronger with a 99.4\% probability 
(K-S test) that
the distribution of mass around MAMBO source locations is different from 
that of random positions in the COSBO survey region.
This signal is dominated by a
group of $\sim 7-8$ of the most significant MAMBO sources close to two
compact mass spikes, which are identified with X-ray bright
over-densities consisting of a total of $\sim 127$ galaxies at $z\approx 0.24$
(Structure \#17 in \citealp{scoville07b})  
and are likely clusters. 
The possible spatial correlation of COSBO sources with
foreground structures, therefore, might be of a somewhat different nature than
that of AzTEC sources. The COSBO region may be
witnessing amplification caused by the two clusters revealed
by the weak-lensing map, while the AzTEC sources are more likely
amplified by galaxies contained within the more
tenuous filamentary large-scale structure, which is so tenuous in the
COSBO field that it provides no significant signal in the sample of MAMBO sources.  

\section{Conclusions}

The central 0.15~deg$^2$ of the AzTEC/COSMOS survey shows a significant 
over-density of bright 1.1~mm-detected SMGs 
when compared to the background population
inferred by other surveys.
We find that this over-density cannot be explained
as sample variance of the blank-field SMG population.
The SMG positions are significantly correlated with the 
$z\lsim$~1.1 optical/IR galaxy density on the sky, which 
is believed to be in the foreground of nearly all AzTEC/COSMOS
SMGs.  
Both the spatial correlation and the AzTEC/COSMOS SMG number counts are consistent
with gravitational amplification of the blank-field 
SMG population. 
The lack of strong correlation to the weak-lensing maps of \citet{massey07}
indicates that this amplification is primarily due to weak-lensing by 
the large-scale structure as opposed to lensing by the compact and 
massive clusters in the field.   
SMGs detected in a different part of the COSMOS field by 
the 1.2~mm COSBO survey 
are also spatially correlated to 
the $z\lsim$~1.1 structure, however, this correlation 
is dominated by two compact structures (likely clusters) 
in the field.
The lack of
significant large-scale structure (i.e. lensing opportunities) 
in the rest of the COSBO survey region
results in COSBO number counts that are 
consistent with the blank-field \citep{bertoldi07} -- 
a strong contrast to the significant SMG over-density and rich foreground structure
found in the nearby AzTEC/COSMOS field.

\section*{Acknowledgements} 
We thank the referee for their thorough reading and helpful comments.
Support for this work was provided in part by NSF grant AST 05-40852 and
a grant from the Korea Science \& Engineering
Foundation (KOSEF) under a cooperative Astrophysical Research Center
of the Structure and Evolution of the Cosmos (ARCSEC).
IA and DHH acknowledge partial support by CONACyT from research grants
39953-F and 39548-F.  

\bibliography{references}

\end{document}